\RequirePackage[mathlines]{lineno}
\documentclass{nature}
\usepackage{graphicx}
\usepackage{verbatim}
\usepackage{color}
\usepackage{float}

\newcommand{\Heebar}{$^4\overline{\rm He}$}
\newcommand{\Hee}{$^4$He}
\newcommand{\He} {$^3$He}
\newcommand{\alphabar}{$\overline{\alpha}$}
\newcommand{\tbar}{$^3\overline{\rm H}$} 
\newcommand{\dbar}{$\overline{\rm d}$}
\newcommand{\hypertbar}{$^3_{\bar{\Lambda}} \overline{\rm H}$}
\newcommand{\Hebar}{$^3\overline{\rm He}$}

\title{Observation of the antimatter helium-4 nucleus  {\footnotesize ( \it{submitted to Nature, under media embargo})}}

\begin{document}

\maketitle


\begin{abstract}
High-energy nuclear collisions create an energy density similar to that of the universe microseconds after the Big Bang\cite{Lemaitre}, and in both cases, matter and antimatter are formed with comparable abundance.  However, the relatively short-lived expansion in nuclear collisions allows antimatter to decouple quickly from matter, and avoid annihilation.  Thus, a high-energy accelerator of heavy nuclei is an efficient means of producing and studying antimatter. The antimatter helium-4 nucleus (\Heebar), also known as the anti-$\alpha$ (\alphabar), consists of two antiprotons and two antineutrons (baryon number $B = - 4$).
It has not been observed previously, although the $\alpha$ particle was identified a century ago by Rutherford and is present in cosmic radiation at the 10\% level\cite{WiebelSooth}.  Antimatter nuclei with $B < - 1$ have been observed only as rare products of interactions at particle accelerators, where the rate of antinucleus production in high-energy collisions decreases by about $1000$ with each additional antinucleon\cite{E864mass, Haidong, Andronic}.  We present the observation of the antimatter helium-4 nucleus, the heaviest observed antinucleus. In total 18 \Heebar~counts were detected at the STAR experiment at RHIC\cite{RHIC} in $10^{9}$ recorded gold on gold (Au+Au) collisions at center-of-mass energies of 200 GeV and 62 GeV per nucleon-nucleon pair. The yield is
consistent with expectations from thermodynamic\cite{PBM} and coalescent nucleosynthesis\cite{sato} models, 
which has implications beyond nuclear physics.
 
\end{abstract}

In 1928, Dirac\cite{Dirac} predicted the existence of negative energy states of electrons based on the application of symmetry principles to quantum mechanics, but only recognized these states to be antimatter after Anderson's\cite{Anderson} discovery of the positron (the antielectron) in cosmic radiation four years later. The predicted antiprotons\cite{Chamberlain} 
and antineutrons\cite{Cork} were observed in 1955, followed by antideuterons 
(\dbar), antitritons (\tbar), and antihelium-3 
(\Hebar) during the following two decades\cite{Zichichi,Lederman, AntiTritium, AntiHe3}.  Recent accelerator and detector advances 
led to the first production of antihydrogen\cite{antiHydrogenCERN} atoms in 1995 and
the discovery of strange antimatter, the antihypertriton (\hypertbar), in 2010 at the Relativistic Heavy Ion Collider (RHIC) at Brookhaven National Laboratory (BNL)\cite{H3Lbar}.

Collisions of relativistic heavy nuclei  
create suitable conditions for producing antinuclei, because large amounts of energy are deposited into 
a more extended volume\cite{Lee} than that achieved in elementary particle collisions.
These nuclear interactions briefly ($\sim10^{-23}$ seconds) produce hot and dense matter 
containing roughly equal numbers of quarks and antiquarks\cite{RHICwhitepaper}, often
interpreted as quark gluon plasma (QGP)\cite{Heinz}. 
In contrast to the Big Bang, nuclear collisions produce 
negligible gravitational attraction and allow the plasma to expand 
rapidly. The hot and dense matter cools down and transitions into a hadron gas, producing nucleons and their 
antiparticles. The production of light antinuclei can be modeled successfully by macroscopic thermodynamics\cite{PBM}, 
which assumes energy equipartition, or by a microscopic coalescence process\cite{BP, sato}, which assumes
uncorrelated probabilities for antinucleons close in position and momentum to become bound. The
high temperature and high antibaryon density of relativistic heavy ion collisions provide a favorable environment
for both production mechanisms. 


\begin{figure}[h!] 
\centering
\makebox[0cm]{\includegraphics[angle=0,width=0.7 \textwidth]{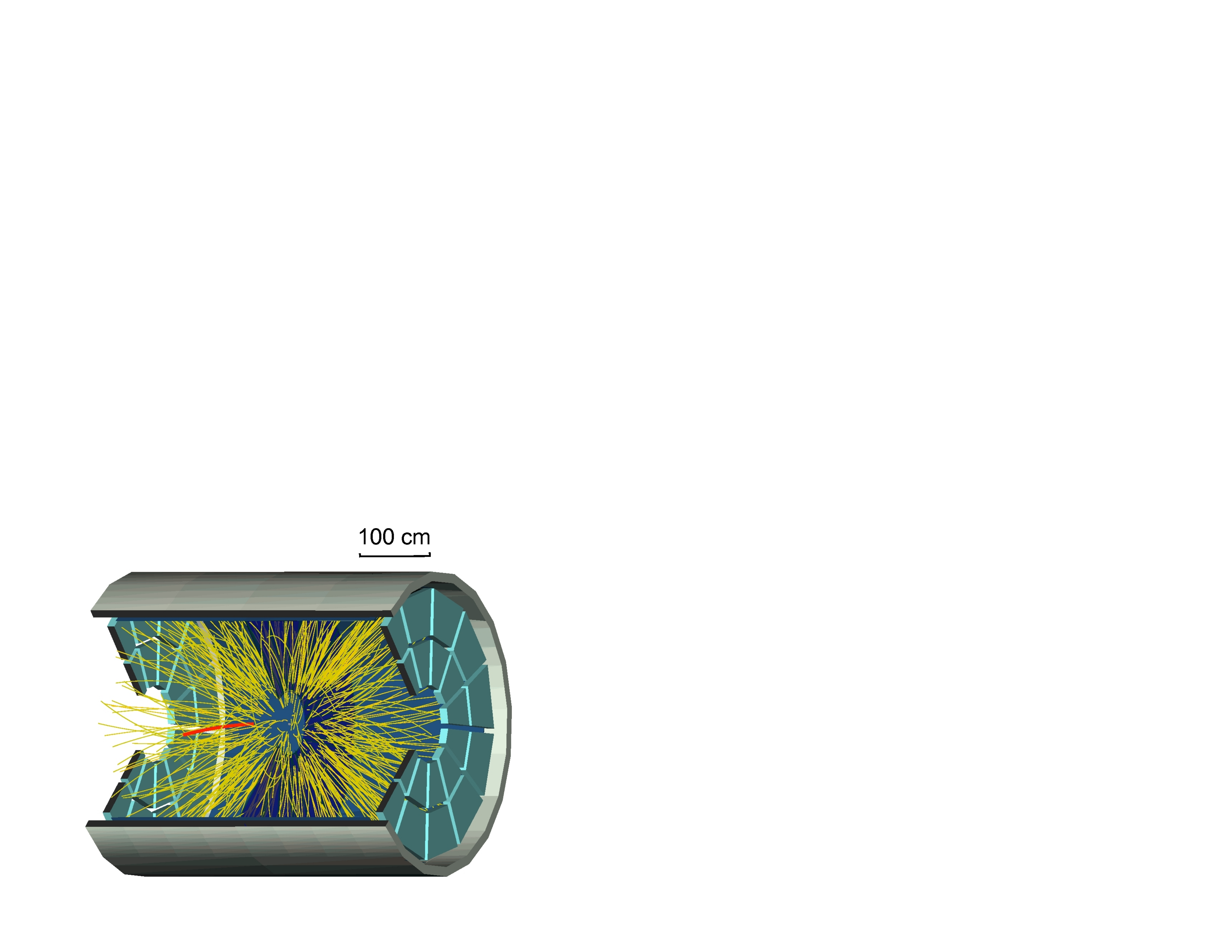}}
\caption{A three-dimensional rendering of the STAR TPC surrounded by the TOF barrel shown as the outermost cylinder. Tracks from an 
event which contains a \Heebar~ are shown, with the \Heebar~track highlighted in bold red.}
\label{fig:alphaInTPC}
\end{figure}        

The central detector used in our measurements of antimatter, 
the Time Projection Chamber (TPC)\cite{TPC} of the STAR experiment 
(Solenoidal Tracker At RHIC), is situated in a solenoidal magnetic field and is 
used for three-dimensional imaging of the ionization trail left along the path of charged particles 
(Fig.~\ref{fig:alphaInTPC}). In addition to the momentum
provided by the track curvature in the magnetic field, 
the detection of \Heebar~particles relies on two key measurements: 
 the mean energy loss per unit track length $\langle dE/dx \rangle$ in the TPC gas, which 
helps distinguish particles with different masses or charges, and the 
time of flight of particles arriving at the time of flight barrel (TOF)\cite{TOF} 
surrounding the TPC.  In general, time of flight provides particle identification in a 
higher momentum range than $\langle dE/dx \rangle$.  The $\langle dE/dx \rangle$ resolution is $7.5\%$ and the timing resolution for TOF is
$95$ picoseconds.

The trigger system at STAR selects collisions of interest for analysis.  
The minimum-bias (MB) trigger selects all particle-producing collisions, regardless of the extent of 
 overlap of the incident nuclei.  A central trigger (CENT) preferentially selects 
head-on collisions, rejecting about $90\%$ of the events acquired using the MB 
trigger.  The sample of $10^9$ Au+Au collisions used in this search is selected based on MB, CENT, and on various specialized triggers.
Preferential selection of events containing tracks with charge $Ze = \pm 2e$ 
(where $e$ is the electron charge) was implemented using a High-Level Trigger (HLT)
for data acquired in 2010. The HLT used computational resources at STAR to perform a 
real-time fast track reconstruction to tag events that had at least one track with 
a $\langle dE/dx \rangle$ value that is larger than a threshold 
set to three standard deviations below the theoretically expected value\cite{Bichsel} for \Hebar~
at the same momentum. The HLT successfully identified $70\%$ of the events where a 
\Heebar~track was present while selecting only 0.4\% of the events for express analyses.


\begin{figure}[H]       
\centering
\makebox[0cm]{\includegraphics[width=0.9 \textwidth]{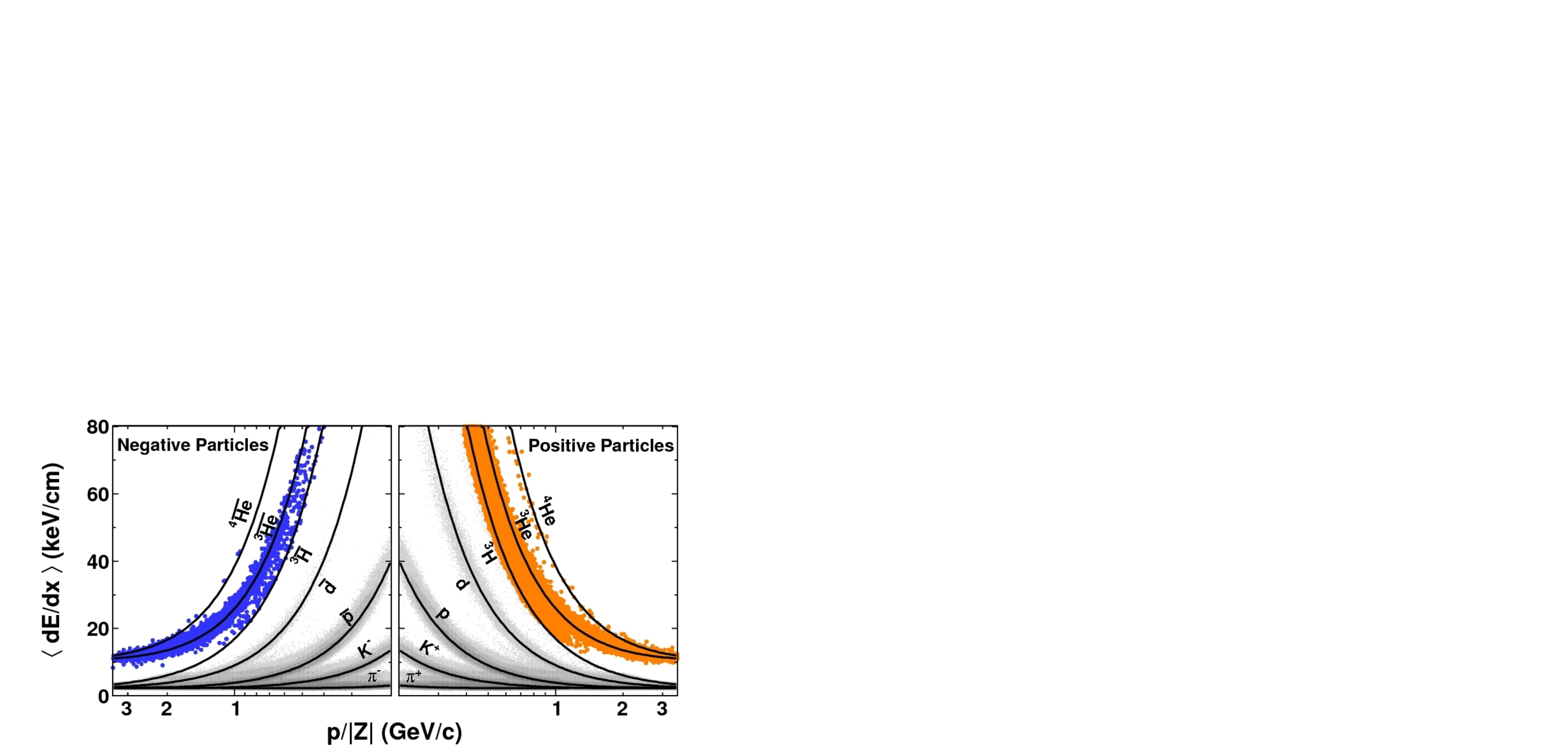}}
\caption{$\langle dE/dx \rangle$ versus $p/|Z|$ for negatively charged particles (left) 
and positively charged particles (right).  The black curves show the 
expected values for each species. The lower edges of the colored bands
correspond to the HLT's online calculation of $3\sigma$ below the $\langle dE/dx \rangle$ band center\cite{Bichsel} for \He. 
}
  \label{fig:dedxvsRigidity}
\end{figure}        

Figure~\ref{fig:dedxvsRigidity} shows $\langle dE/dx \rangle$ 
versus the magnitude of magnetic rigidity, $p/|Z|$, where $p$ is momentum.  A distinct band of 
positive particles centered around the expected value\cite{Bichsel} for \Hee~particles indicates that 
the detector is well-calibrated.   
On the left side of Fig.~\ref{fig:dedxvsRigidity}, where $p/|Z|$ is less than 1.4 GeV$/c$,
four negative particles are particularly well separated from the \Hebar~band and are located within the 
expected band for \Heebar.
Above 1.75 GeV$/c$, $\langle dE/dx \rangle$ values of \Hebar ~ and \Heebar ~ merge and 
the TOF system is needed to separate these two species.  

\begin{figure}[H]       
\centering
\makebox[0cm]{\includegraphics[width=0.6 \textwidth]{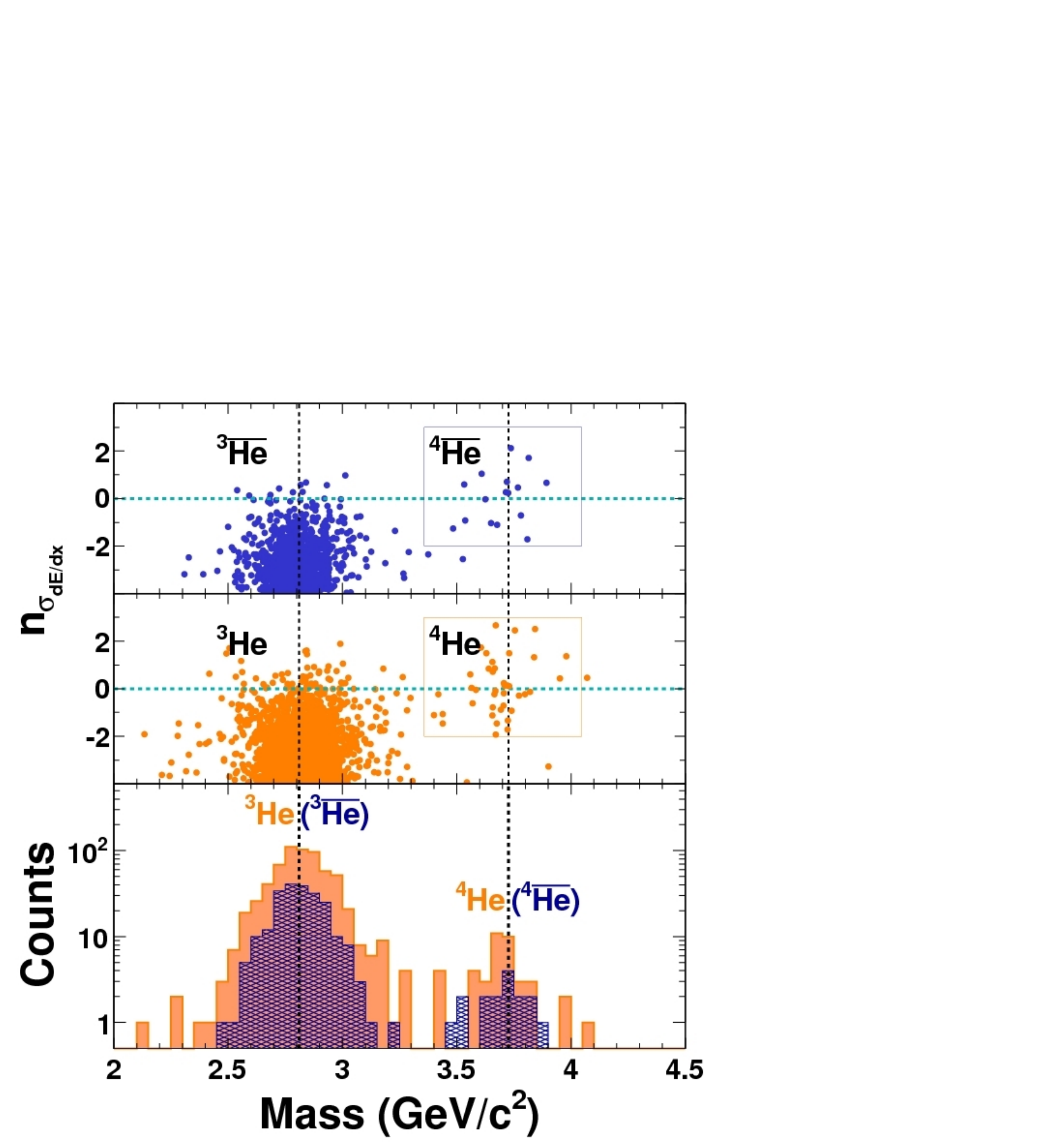}}
\caption{The top two panels show the $\langle dE/dx \rangle$ 
in units of multiples of $\sigma_{dE/dx}$, $n_{\sigma_{dE/dx}}$,
of negatively charged particles 
(first panel) and positively charged particles (second panel) as a function of mass measured by the TOF system. The masses of \He~(\Hebar) and  \Hee~(\Heebar)
are indicated by the vertical lines at 2.81 GeV/$c^2$ and 3.73 GeV/$c^2$, respectively.
The horizontal line marks the position of zero 
deviation from the expected value of $\langle dE/dx \rangle$ ($n_{\sigma_{dE/dx}}=0$) 
for \Hee~(\Heebar). The rectangular boxes highlight areas for \Hee~(\Heebar)~selections : 
$-2 < n_{\sigma_{dE/dx}}  < 3$ and $3.35 \, \mathrm{GeV}/c^2 < \mathrm{mass} 
< 4.04\, \mathrm{GeV}/c^2$ (corresponding to a $\pm 3 \sigma$ window in mass).
The bottom panel shows a  projection of entries in the upper two panels onto the mass axis 
for particles in the window of $-2 < \sigma_{dE/dx} < 3$.  
The combined measurements of energy loss and the time of flight allow a
clean identification to be made in a sample of $0.5\times10^{12}$ tracks from $10^9$ Au+Au collisions.
}
  \label{fig:m2VsNSigma}
\end{figure}

The top two panels of Fig.~\ref{fig:m2VsNSigma} shows the
$\langle dE/dx \rangle$ (in units of multiples of $\sigma_{dE/dx}$, $n_{\sigma_{dE/dx}}$)
 versus  calculated mass $m = (p/c)\sqrt{(t^2c^2/L^2 -1)}$,
where $\sigma_{dE/dx}$ is the rms width of the $\langle dE/dx \rangle$ distribution for \Hee 
~or \Heebar, $t$ and $L$ are the time of flight and path 
length, respectively, and $c$ is the speed of light. The 
first (second) panel shows negatively (positively) charged particles.  In both panels, majority
species are \He~ and \Hebar. In the second panel, the 
\Hee~particles cluster around $n_{\sigma_{dE/dx}} = 0$ and mass $= 3.73$ 
GeV$/c^2$, the appropriate mass for \Hee.  A similar but smaller cluster of particles can be 
found in the first panel for \Heebar.  The bottom panel shows the projection onto the mass axis of the top two 
panels for particles with $n_{\sigma_{dE/dx}}$ of -2 to 3.  There is clear separation  
between \Hebar~and \Heebar~ mass peaks.  Eighteen counts for \Heebar~are observed. Of those,
sixteen are from collisions recorded in 2010. Two counts\cite{Zhou} identified by $\langle dE/dx \rangle$
alone from data recorded in 2007 are not included in this figure, because the STAR
TOF was not installed at that time.

To evaluate the background in \Heebar~due to \Hebar~contamination, 
we simulate the \Hebar~mass distribution with momenta and path lengths, as well as the
expected time of flight from \Hebar~particles with timing resolution derived from the same data sample.
The contamination from misidentifying \Hebar~as \Heebar~is estimated by integrating over the region of the \Heebar~selection. We
estimate that the background contributes 1.4 (0.05) counts of the 15 (1) total counts from 
Au+Au collisions at 200 (62) GeV recorded in 2010. 

\begin{figure}[H] 
\centering
\makebox[0cm]{\includegraphics[width=0.6 \textwidth]{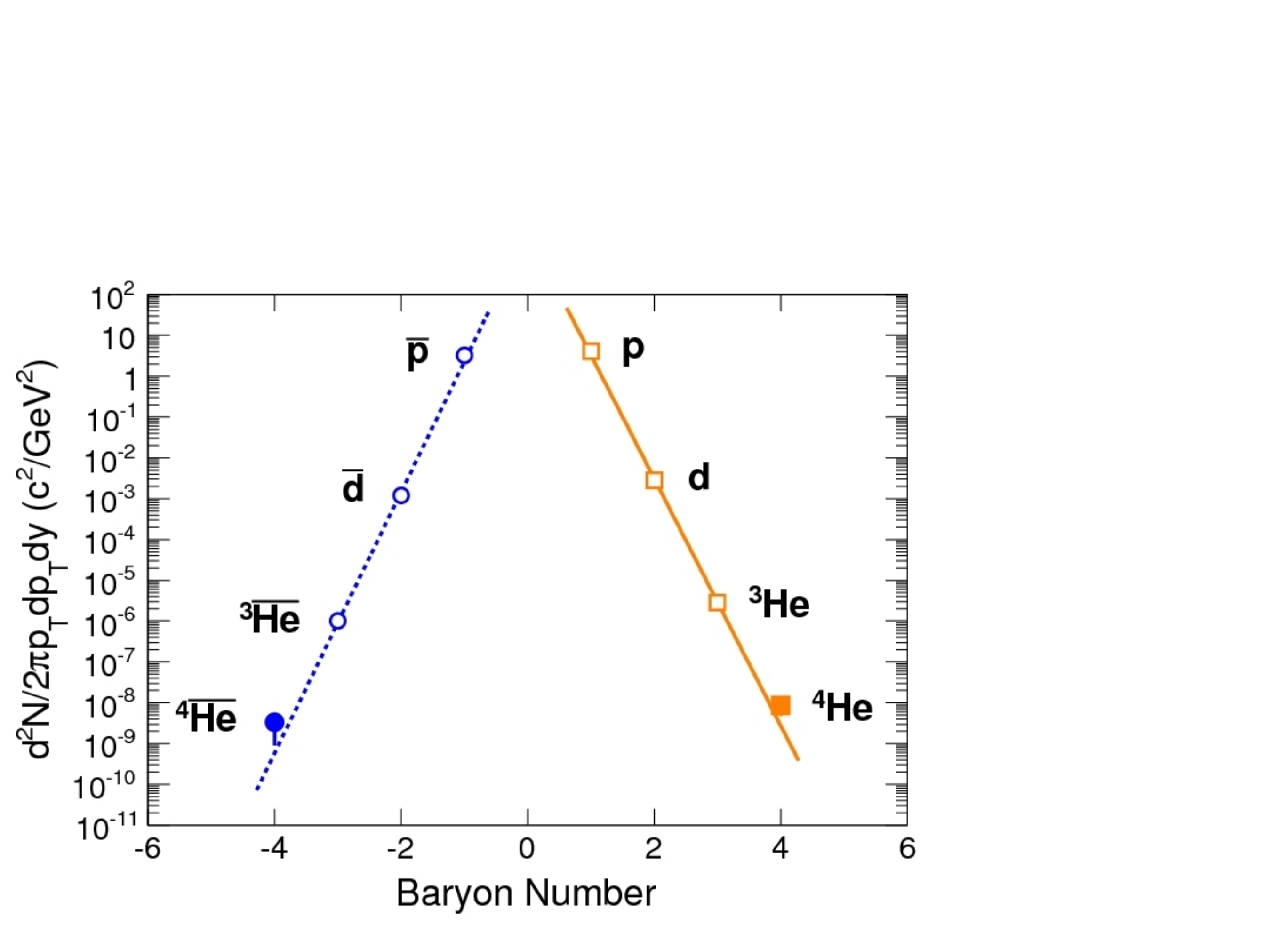}}
\caption{Differential invariant yields as a function of baryon number $B$, evaluated at $p_{T}/|B| = 0.875\,\, \mathrm{GeV}/c$, in central 200 GeV 
Au+Au collisions.  Yields for (anti)tritons ($^3$H and \tbar) lie close to the 
positions for \He~and \Hebar, but are not included here because of poorer 
identification of (anti)tritons.  The lines represent fits with the exponential formula $\propto e^{-r|B|}$ for 
positive and negative particles separately, where $r$ is
the production reduction factor.  Analysis details of yields 
other than \Hee~(\Heebar)~have been presented elsewhere\cite{He3Yield ,Haidong}.  Errors are statistical 
only. Systematic errors are smaller than the symbol size, and are not plotted.
}
  \label{fig:yields}
\end{figure}

The observed counts are used to calculate the antimatter yield with appropriate normalization (the differential invariant yield)
in order to compare to the theoretical expectation.
Detector acceptance, efficiency, and antimatter annihilation with the detector 
material are taken into account when computing yields. 
Various uncertainties related to tracking in the TPC, 
matching in the TOF, and triggering in the HLT are cancelled when 
the yield ratios of \Hee/\He~and \Heebar/\Hebar~ 
are calculated. The ratios are 
\Hee/\He $~=(3.0\pm 1.3(\mathrm{stat}) _{-0.3}^{+0.5}(\mathrm{sys}))\times 10^{-3}$ and 
\Heebar/\Hebar$~=(3.2\pm 2.3(\mathrm{stat}) _{-0.2}^{+0.7}(\mathrm{sys}))\times 10^{-3}$ for central Au+Au collisions at 200 GeV.
The ratios were obtained in a window $40^\circ < \theta < 140^\circ$, where polar angle, $\theta$, is the angle between the 
particle's momentum vector and the beam axis (these $\theta$ limits correspond to limits of $-1$ to $1$ in a related quantity, pseudorapidity), and in a $p_T$ per baryon window centered at $p_T/|B|$ = 0.875 GeV/$c$ with a width of 0.25 GeV/$c$, where $p_T$ is the projection of the momentum vector on the plane that is transverse to the beam axis.
Ratios calculated by a Blastwave model\cite{TBW} for the $p_T/|B|$ window mentioned above and
for the whole range of $p_T/|B|$ differ by only 1\%. 
The differential yields ($d^2N/(p_T dp_T dy)$) for \Hee~(\Heebar) are obtained by
multiplying the ratio of \Hee/\He~(\Heebar/\Hebar) with the \He~(\Hebar) yields\cite{He3Yield}. 
The systematic uncertainties consist of background (-6\% for both ratios), 
feed-down from (anti-)hypertritons (18\% for both \He~and \Hebar), 
knock-outs from beam-material interactions (-5\% for the ratio \Hee/\He~only) and 
absorption (4\% for the ratio \Heebar/\Hebar~only).
Figure~\ref{fig:yields} shows the exponential\cite{E864mass} invariant yields 
versus baryon number in 200 GeV central Au+Au collisions.
Empirically, the production rate reduces by a factor of 
$1.6_{-0.6}^{+1.0} \times 10^{3}$ ($1.1_{-0.2}^{+0.3} \times 10^{3}$)
for each additional antinucleon (nucleon) added to the antinucleus 
(nucleus). This general trend is expected from coalescent nucleosynthesis models\cite{sato}, 
originally developed to describe production of antideuterons\cite{BP},
and as well as from thermodynamic models\cite{PBM}.

In a microscopic picture, a light 
nucleus emerging from a relativistic heavy-ion collision is produced 
during the last stage of the collision process. The quantum wave functions of 
the constituent nucleons, if close enough in momentum and coordinate space, will 
overlap to produce the nucleus. The production rate for a nucleus with baryon number 
$B$ is proportional to the nucleon density in momentum and coordinate space, raised to the power of $|B|$, 
and therefore exhibits exponential behaviour as a function of $B$. 
Alternatively, in a thermodynamic 
model, a nucleus is regarded as an object with energy $E \sim |B| m_N$, where 
$m_N$ is the nucleon mass, and the production rate 
is determined by the Boltzmann factor $\exp{(-E/T)}$, where $T$ is the 
temperature\cite{PBM,E864mass}. This model also produces an exponential 
yield. A more rigorous calculation\cite{Andronic} can provide a good fit to the available particle yields and
predicts the ratios integrated over $p_T$ to be 
\Hee/\He $~=3.1\times 10^{-3}$ and \Heebar/\Hebar~
$=2.4\times 10^{-3}$, consistent with our measurements. 
The considerations outlined above offer a good estimate for the production rate of 
even heavier antinuclei. For example, the yield of the stable antimatter nucleus next in line ($B = - 6$)
is predicted to be down by a factor of $2.6\times10^6$ compared to \Heebar~and is 
beyond the reach of current accelerator technology.

A potentially more 
copious production mechanism for heavier antimatter is by the direct excitation of complex 
nuclear structures from the vacuum\cite{wgreiner}. A deviation from the usual 
rate reduction with increasing mass would be an indication of a radically new 
production mechanism\cite{PBM}. On the other hand, going beyond nuclear physics, 
the sensitivity of current and planned space-based charged particle detectors is below what would be
needed to observe antihelium produced by nuclear interactions in the cosmos, and consequently,
any observation of antihelium or even heavier antinuclei in space would indicate the existence of
a large amount of antimatter elsewhere in the universe. In particular, finding \Heebar~
in the cosmos is one of the major motivations for space detectors 
such as the Alpha Magnetic Spectrometer\cite{AMS}.
We have shown that \Heebar~exists, and have measured its rate of production in nuclear interactions, providing
a point of reference for possible future observations in cosmic radiation.
Barring one of those dramatic 
discoveries mentioned above or new breakthrough in the accelerator technology,
it is likely that the \Heebar~will remain the heaviest stable 
antimatter nucleus observed for the foreseeable future.

\begin{addendum}
 \item We thank the RHIC Operations Group and RACF at BNL, the NERSC Center at LBNL and the Open Science Grid consortium for providing resources and support. This work was supported in part by the Offices of NP and HEP within the U.S. DOE Office of Science, the U.S. NSF, the Sloan Foundation, the DFG cluster of excellence `Origin and Structure of the Universe'of Germany, CNRS/IN2P3, FAPESP CNPq of Brazil, Ministry of Ed. and Sci. of the Russian Federation, NNSFC, CAS, MoST, and MoE of China, GA and MSMT of the Czech Republic, FOM and NWO of the Netherlands, DAE, DST, and CSIR of India, Polish Ministry of Sci. and Higher Ed., Korea Research Foundation, Ministry of Sci., Ed. and Sports of the Rep. Of Croatia, and RosAtom of Russia.
\end{addendum}

\author{H.~Agakishiev$^{17}$, M.~M.~Aggarwal$^{29}$, Z.~Ahammed$^{21}$, 
A.~V.~Alakhverdyants$^{17}$, I.~Alekseev~~$^{15}$, J.~Alford$^{18}$, 
B.~D.~Anderson$^{18}$, C.~D.~Anson$^{27}$, D.~Arkhipkin$^{2}$, 
G.~S.~Averichev$^{17}$
% ...
\\
\\(STAR Collaboration) 
\\
\normalsize{$^{1}$Argonne National Laboratory, Argonne, Illinois 60439, USA}\\ 
\normalsize{$^{2}$Brookhaven National Laboratory, Upton, New York 11973, USA}\\ 
\normalsize{$^{3}$University of California, Berkeley, California 94720, USA}\\ 
\normalsize{$^{4}$University of California, Davis, California 95616, USA}\\ 
\normalsize{$^{5}$University of California, Los Angeles, California 90095, USA}\\ 
\normalsize{$^{6}$Universidade Estadual de Campinas, Sao Paulo, Brazil}\\ 
\normalsize{$^{7}$University of Illinois at Chicago, Chicago, Illinois 60607, USA}\\ 
\normalsize{$^{8}$Creighton University, Omaha, Nebraska 68178, USA}\\ 
\normalsize{$^{9}$Czech Technical University in Prague, FNSPE, Prague, 115 19, Czech Republic}\\ 
\normalsize{$^{10}$Nuclear Physics Institute AS CR, 250 68 \v{R}e\v{z}/Prague, Czech Republic}\\ 
\normalsize{$^{11}$University of Frankfurt, Frankfurt, Germany}\\ 
\normalsize{$^{12}$Institute of Physics, Bhubaneswar 751005, India}\\ 
\normalsize{$^{13}$Indian Institute of Technology, Mumbai, India}\\ 
\normalsize{$^{14}$Indiana University, Bloomington, Indiana 47408, USA}\\ 
\normalsize{$^{15}$Alikhanov Institute for Theoretical and Experimental Physics, Moscow, Russia}\\ 
\normalsize{$^{16}$University of Jammu, Jammu 180001, India}\\ 
\normalsize{$^{17}$Joint Institute for Nuclear Research, Dubna, 141 980, Russia}\\ 
\normalsize{$^{18}$Kent State University, Kent, Ohio 44242, USA}\\ 
\normalsize{$^{19}$University of Kentucky, Lexington, Kentucky, 40506-0055, USA}\\ 
\normalsize{$^{20}$Institute of Modern Physics, Lanzhou, China}\\ 
\normalsize{$^{21}$Lawrence Berkeley National Laboratory, Berkeley, California 94720, USA}\\ 
\normalsize{$^{22}$Massachusetts Institute of Technology, Cambridge, MA 02139-4307, USA}\\ 
\normalsize{$^{23}$Max-Planck-Institut f\"ur Physik, Munich, Germany}\\ 
\normalsize{$^{24}$Michigan State University, East Lansing, Michigan 48824, USA}\\ 
\normalsize{$^{25}$Moscow Engineering Physics Institute, Moscow Russia}\\ 
\normalsize{$^{26}$NIKHEF and Utrecht University, Amsterdam, The Netherlands}\\ 
\normalsize{$^{27}$Ohio State University, Columbus, Ohio 43210, USA}\\ 
\normalsize{$^{28}$Old Dominion University, Norfolk, VA, 23529, USA}\\ 
\normalsize{$^{29}$Panjab University, Chandigarh 160014, India}\\ 
\normalsize{$^{30}$Pennsylvania State University, University Park, Pennsylvania 16802, USA}\\ 
\normalsize{$^{31}$Institute of High Energy Physics, Protvino, Russia}\\ 
\normalsize{$^{32}$Purdue University, West Lafayette, Indiana 47907, USA}\\ 
\normalsize{$^{33}$Pusan National University, Pusan, Republic of Korea}\\ 
\normalsize{$^{34}$University of Rajasthan, Jaipur 302004, India}\\ 
\normalsize{$^{35}$Rice University, Houston, Texas 77251, USA}\\ 
\normalsize{$^{36}$Universidade de Sao Paulo, Sao Paulo, Brazil}\\ 
\normalsize{$^{37}$University of Science \& Technology of China, Hefei 230026, China}\\ 
\normalsize{$^{38}$Shandong University, Jinan, Shandong 250100, China}\\ 
\normalsize{$^{39}$Shanghai Institute of Applied Physics, Shanghai 201800, China}\\ 
\normalsize{$^{40}$SUBATECH, Nantes, France}\\ 
\normalsize{$^{41}$Texas A\&M University, College Station, Texas 77843, USA}\\ 
\normalsize{$^{42}$University of Texas, Austin, Texas 78712, USA}\\ 
\normalsize{$^{43}$University of Houston, Houston, TX, 77204, USA}\\ 
\normalsize{$^{44}$Tsinghua University, Beijing 100084, China}\\ 
\normalsize{$^{45}$United States Naval Academy, Annapolis, MD 21402, USA}\\ 
\normalsize{$^{46}$Valparaiso University, Valparaiso, Indiana 46383, USA}\\ 
\normalsize{$^{47}$Variable Energy Cyclotron Centre, Kolkata 700064, India}\\ 
\normalsize{$^{48}$Warsaw University of Technology, Warsaw, Poland}\\ 
\normalsize{$^{49}$University of Washington, Seattle, Washington 98195, USA}\\ 
\normalsize{$^{50}$Wayne State University, Detroit, Michigan 48201, USA}\\ 
\normalsize{$^{51}$Institute of Particle Physics, CCNU (HZNU), Wuhan 430079, China}\\ 
\normalsize{$^{52}$Yale University, New Haven, Connecticut 06520, USA}\\ 
\normalsize{$^{53}$University of Zagreb, Zagreb, HR-10002, Croatia}\\ 
}

\pagebreak


\end{document}